# Photoconductivité et photoémission de diamant(s) sous irradiation XUV femtoseconde

J. Gaudin [1], G. Geoffroy [1], S. Guizard [1], S. Esnouf [1], V. Olevano [1], G. Petite [1], S.M. Klimentov [2], P.A. Pivovarov [2], S.V. Garnov [2], B. Carre [3], P. Martin [4] et A. Belsky [4]

*(1) Laboratoire des Solides Irradiés, UMR 7642, CEA/DSM, CNRS/SPM et Ecole Polytechnique, 91128 Palaiseau Cedex, France*
*(2) General Physics Institute of the RAS , 38 Vavilov St., 119191, Moscou, Russie*
*(3) Service Photons Atomes et Molecules, CEA/DSM, Cea Saclay, F-91191, Gif sur Yvette*
*(4) CELIA (UMR 5107), Université Bordeaux I, 351 cours de la libération, F-33405 Talence*

Résumé : Nous décrivons une étude des propriétés de photoconductivité (PC) induite dans différents types de diamants (monocristaux de type IIa et couches CVD) par des impulsions femtosecondes XUV (jusqu'à l'harmonique 19 du laser titane/saphir). En complément de ces études, les spectres de photoémission de ces échantillons ont aussi été étudiés (harmoniques 13 à 27).
En fonction de l'ordre de l'harmonique, on constate que le signal de PC augmente tout d'abord (harmoniques 9 à 13) puis diminue au delà. Si l'augmentation s'interprète aisément comme résultant de phénomènes de multiplication par collisions inélastiques, la diminution ultérieure n'a pas pour le moment d'explication. Les mesures de spectre de photoémission suggèrent un effet important de la relaxation par émission de plasmons. Enfin, nous avons réalisé le premier calcul ab-initio de la durée de vie des porteurs tenant compte des interactions électron-électron, à l'aide d'une approche de théorie quantique à plusieurs corps de type GW. Au voisinage du gap, on observe un comportement proche de celui d'un liquide de Fermi. A plus haute énergie on observe des déviations à ce comportement, provenant d'effets de structure de bande d'une part, et d'excitations de plasmons d'autre part.

Dans le cadre des études sur l'effet des rayonnements ionisants sur les isolants, il est particulièrement important de bien comprendre le comportement des porteurs excités dans ces matériaux, ce qui reste un problème largement ouvert. L'étude présentée ci dessous participe de cet objectif, en s'appuyant sur les remarques suivantes :
- les cinétiques relatives à ces porteurs excités sont extrêmement rapides, et l'utilisation de sources laser ou XUV femtoseconde est de ce point de vue pertinente.
- le diamant, dans ses formes variées, représente un cas type d'isolant à grande bande interdite purement covalent, et sa structure simple permet d'en modéliser correctement un grand nombre de propriétés. Par ailleurs ce matériau fait l'objet d'applications diverses, telles par exemple que la détection de particules, pour lesquelles les propriétés électroniques sont de première importance.

Nous rendons compte ci dessous des résultats obtenus dans deux séries d'expériences portant sur la photoconductivité transitoire induite dans le diamant par des impulsions laser XUV femtoseconde ainsi que la spectroscopie de photoémission du diamant utilisant des impulsions semblables. Par ailleurs, nous présenterons les premiers calculs ab-initio effectués sur la durée de vie des porteurs photoinjectés dans le diamant.

Les échantillons étudiés sont de deux types : un échantillon polycristallin obtenu par CVD et un monocristal de diamant naturel de type IIa. Préalablement aux expériences, ces échantillons ont été caractérisés par RPE (bande X et bande Q), ce qui a permis de détecter
- dans l'échantillon CVD, le triplet hyperfin correspondant au centre P1, associé à la présence d'azote substitutionnel dans l'échantillon, et un signal large, traditionnellement attribué dans la littérature à la présence de liaisons pendantes.
- dans le monocristal IIa, uniquement le signal correspondant aux liaisons pendantes, et ceci au niveau de la limite de détection du spectromètre bande Q. Le signal de P1 n'a pu être détecté dans cet échantillon qui apparaît comme exceptionnellement pur.

Les sources de rayonnement femtoseconde utilisées dans ce travail sont obtenues par génération d'harmoniques élevées d'un laser au saphir dopé titane émettant à 800nm (hv=1.55eV). Nous avons utilisé pour l'expérience de photoconductivité les harmoniques 7 à 19, générées par la source LUCA (DRECAM/CEA Saclay), et pour l'expérience de spectroscopie de photoémission, les harmoniques 13 à 29 générées par le laser du CELIA (Bordeaux-I).

Le figure 1 décrit le principe de la mesure de photoconductivité pulsée. L'échantillon est placé entre deux électrodes métalliques, sans contact avec celles-ci. Simultanément avec l'impulsion XUV, une impulsion de haute tension (de l'ordre du kV) est appliquée à l'électrode supérieure. L'impulsions XUV provoque l'excitation de porteurs libres (électrons et trous) dans l'échantillon. Leur déplacement sous l'influence du champ appliqué induit un courant de polarisation, qui est transmis par effet capacitif aux électrodes. Une impulsion de courant est donc mesurée sur l'électrode basse tension, qui reflète le nombre de porteurs injectés.

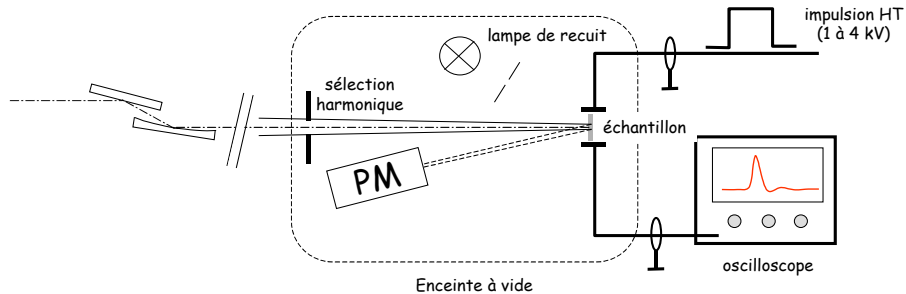

Figure 1 : dispositif expérimental de mesure de la photoconductivité induite par impulsions XUV

La mesure de l'intensité du rayonnement incident est effectuée sur la réflexion de ce dernier sur l'échantillon, à l'aide d'un photomultiplicateur (ce qui pose des problèmes d'étalonnage, qui seront évoqués ci dessous).

L'amplitude de l'impulsion de photoconductivité détectée à l'oscilloscope ($U_{PC}$) s'écrit [1]

$$U_{PC} = \alpha \frac{\mu e U_0}{L^2 R} N \qquad (1)$$

où $\alpha$, $L$ et $R$ sont des coefficients dépendant de la géométrie expérimentale, $\mu$ est la mobilité des porteurs, $U_0$ la tension appliquée et $N$ le nombre de porteurs injectés (un seul type considéré ici). La mesure, toutes les autres quantités étant fixées, de la variation de $U_{PC}$ en fonction de l'ordre de l'harmonique utilisée ( pulsation $\omega$) produit donc en principe une mesure (discrète) du rendement de photoinjection. Cependant, il faut remarquer ici que la totalité de photons pénétrant dans l'échantillon sont absorbés, et qu'en principe chacun induit la création d'une paire e-h : à nombre de photons absorbés constant, on attend un signal indépendant de l'ordre de l'harmonique, et une dépendance $U_{PC}(\omega)$ traduirait donc en principe l'existence d'un phénomène de « multiplication des porteurs » qui est l'objet de la présente étude. Avant de détailler ce point, signalons que l'eq. (1) est valable à condition que le temps de réponse de l'appareillage (de l'ordre de 1.5 ns) soit plus grand que les autres constantes cinétiques du problème, et il est donc important de vérifier que la forme du signal est stable en fonction de la densité d'excitation (ce qui est le cas dans nos expériences). Signalons un processus cinétique important (en plus de la durée de l'impulsion excitatrice, et du temps de piégeage des porteurs) propre à ce type d'expérience : la limitation du signal par formation de charge d'espace. Sous l'influence de l'impulsion incidente, on crée une distribution de paire e-h qui se séparent sous l'influence du champ électrique appliqué. Ce faisant, la distribution de charge produite crée progressivement un champ interne écrantant le champ appliqué, et ce jusqu'à ce que la résultante des deux champs soit nulle (le courant de polarisation s'interrompt alors). Il est important que cette distribution de charge induite soit revenue à l'équilibre avant la mesure suivante, ce qui se fait, quand la tension appliquée est pulsée, dès que celle ci s'interrompt, sous l'influence du champ interne de charge d'espace. On peut accélérer ce processus en utilisant, pour dépiéger les porteurs, une irradiation visible (« lampe de recuit » sur la figure 1), qui provoque aussi un échauffement de l'échantillon.. La fréquence d'acquisition doit donc être adaptée aux limites imposées par ce phénomène, le temps minium entre deux mesures variant selon le type d'échantillon (CVD ou monocristaux), probablement à cause des densités très différentes de sites de piégeage des charges.

Avant de passer à la description des résultats expérimentaux, disons quelques mots du processus de multiplication des porteurs dont le principe est illustré sur la figure 2 (qui présente une situation ultra-simplifiée : bande de conduction parabolique, bande de valence « plate », c'est à dire trous infiniment lourds). Un électron injecté à une énergie suffisante dans la bande de conduction peut, par collision avec un électron de la bande de valence créer une seconde paire e-h (ce qui revient à doubler dans notre expérience le signal de photoconductivité). L'énergie seuil d'excitation nécessaire pour que ceci se produise dépend de la structure de bande. Dans la situation de la figure 2 – gap direct, trous infiniment lourds - elle est égale (exprimée en énergie de photon harmonique) à deux fois l'énergie de bande interdite ($E_G$). Avec une masse équivalente des électrons et des trous, elle passe à 4 $E_G$. Il n'y a par contre pas de prédiction simple pour le cas d'un matériau à gap indirect comme le diamant, pour lequel la structure de bande est par ailleurs passablement compliquée [2]. Des mesures sur des matériaux semblables appliqués à l'électronique (Si et Ge) donnent des résultats se situant entre 3 et 4 $E_G$. Une information expérimentale sur ce point est donc un élément important, dans la mesure où elle est un élément essentiel de l'évaluation de la sensibilité du diamant comme détecteur de rayonnement UVX (et nous informe sur les mécanismes sous-jacents). Sans reporter les détails de la structure de bande [2], nous montrons sur la figure 3 les énergies d'excitation possibles dans le cas du diamant. Notons que celui ci possède deux bandes de conduction, séparées par des gap directs importants (mais le gap réel est beaucoup plus faible : 1 eV ou moins). Du point de vue de l'excitation par les harmoniques, on constate que l'harmonique 13 ($E_P$= 20.2 eV) ne permet pas d'exciter dans la deuxième bande de conduction, ce qui est par contre possible avec l'harmonique 15 ($E_P$= 23.6 eV). Nous avons réalisé un calcul LDA de cette même structure de bande qui confirme essentiellement les résultats de [2], tout en prédisant moins bien le gap, ce qui est un défaut connu de ce type de théorie.

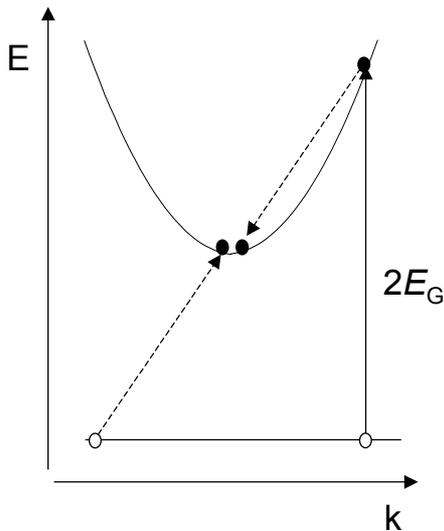

Figure 2 : principe du mécanisme de multiplication des porteurs par excitation électronique secondaires. La masse du trou est supposée infinie

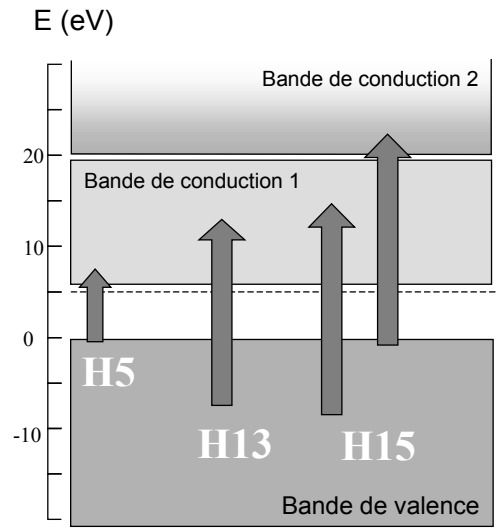

Figure 3 : bandes d'énergies permises dans le diamant (d'après [2]) et excitations possibles avec différentes harmoniques

L'expérience consiste à enregistrer un certain nombre de caractéristiques $U_{PC}(U_{PM}) - U_{PM}$ : signal PM, proportionnel à l'intensité de l'harmonique - et d'en tirer, en se plaçant à intensité harmonique constante un spectre d'excitation (échantillonné par les fréquences des harmoniques disponibles) qui, rappelons le, reflète la totalité des porteurs résultant de l'excitation, et non la probabilité d'absorption initiale du rayonnement. Par « à intensité harmonique constante », il faut entendre « à nombre égal de photons absorbés dans le matériau », ce qui suppose de tenir compte de la sensibilité du spectrale PM (donnée par le constructeur) mais aussi de la réflectivité de l'échantillon. Celle ci peut être trouvée dans la littérature [3,4], mais on constate immédiatement des variations significatives selon les auteurs. Nous avons donc procédé à une calibration de cette réflectivité dans les conditions de notre expérience.

La figure 4 présente de tels spectres d'excitation obtenus sur deux échantillons différents – (a) CVD, montrant le résultats de deux expériences effectuées à quelques mois de distance, la différence d'amplitude du signal pouvant s'expliquer par des configurations légèrement différentes du positionnement électrodes/échantillons, et (b) monocristal IIa – et compare les résultats obtenus avec notre calibration à ceux obtenus en utilisant celle de Walker *et al.* [4]. On constate la aussi une différence d'amplitude des effets mesurés, mais on constate aussi que quelle que soit la calibration, quel que soit l'échantillon, et les détails de la configuration expérimentale, on observe clairement une croissance du signal de l'harmonique 9 à l'harmonique 13, suivie d'une décroissance des harmoniques 13 à 19. Sur la figure 4,b) on a représenté l'évolution du nombre de paires créées par photon absorbé dans l'hypothèse d'un seuil à 2 $E_G$, et on constate immédiatement que l'argument purement énergétique pour prévoir l'évolution du rendement de multiplication des porteurs ne rend absolument pas compte de l'expérience. Il faut évidemment envisager la possibilité d'artefacts, les seuls réellement sérieux étant la compétition entre la photoconductivité et la photoémission (en effet, à cause de la faible pénétration du rayonnement, au maximum 80 nm, certain électrons sont émis et peuvent donc atténuer le signal de PC, les précautions nécessaires ayant été prises pour qu'ils ne puissent atteindre les électrodes), mais celle ci décroît avec l'ordre de l'harmonique [5], et ne peut donc expliquer la chute du signal au delà de l'harmonique 13, et la possibilité d'une contribution de surface, qui n'a pas de raison d'être résonnante et montre généralement un profil temporel très différent.

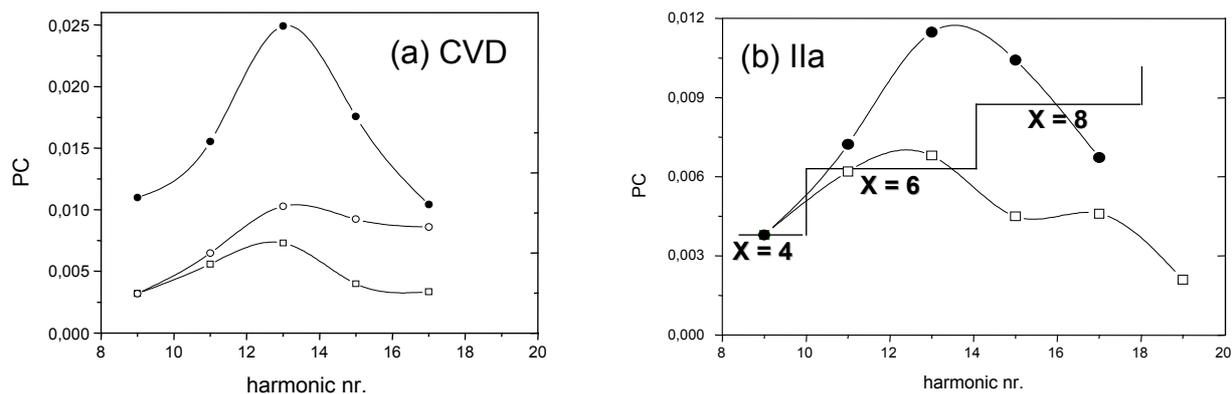

Figure 4 : spectre d'excitation d'échantillons (a) CVD et (b) monocristaux IIa. (a) et (b)   : calibration de /6/, •ou o : notre calibration (les symboles ouverts et fermés correspondent à deux expériences réalisées sur le même échantillon à plusieurs mois de distance). X : nombre de paires créées par photon absorbé dans l'hypothèse d'un seuil à 2 $E_G$.

Nous ne sommes pas en mesure pour le moment de proposer une explication définitive au comportement observé, mais nous pouvons déjà faire les remarques suivantes :
- la première partie de la courbe (croissance de l'harmonique 9 à l'harmonique 13) peut en principe se comprendre sur la base du processus de multiplication des porteurs : l'énergie cinétique disponible dans la bande de conduction 1 dépasse 10 eV, ce qui est en principe suffisant pour exciter une paire e-h supplémentaire.
- on doit noter la coïncidence entre le changement de comportement à partir de l'harmonique 13, et le fait que l'on passe ici le seuil d'excitation de la deuxième bande de conduction. Ceci oriente la recherche d'une explication vers un effet de structure de bande. Un renseignement extrêmement utile dans ce sens serait un calcul du spectre de perte d'énergie électronique, en fonction de l'énergie de porteur, ce qui est en principe possible, mais reste à faire.

Pour la recherche d'une explication de ce type, il est évident que la mesure des spectres d'énergie de photoélectrons peut apporter des renseignement précieux. Nous avons donc réalisé cette mesure sur l'échantillon monocristallin. Pour cela, nous avons utilisé un dispositif classique (spectromètre CLAM IV de type hémisphérique) installé sur la ligne « harmoniques » du laser du CELIA. La difficulté dans une expérience de ce type sur des échantillons isolants tient à la nécessité de se débarrasser des effets de charge de l'échantillon, ce qui est obtenu en chauffant celui ci à une température de 300°C. Cette pratique permet en général de stabiliser le potentiel de surface de l'échantillon.

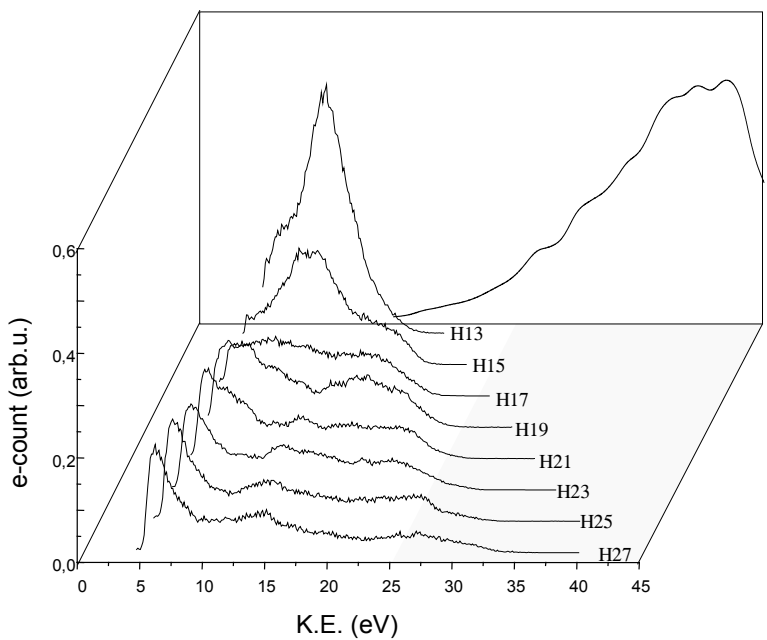

Figure 5 : spectres de photoémission (UPS) enregistrés sur le monocristal IIa avec les harmoniques 13 à 27.

La figure 5 représente une série de spectres obtenus de la sorte pour les harmoniques 13 à 27. Les spectres on été recalés en énergie (des effets classiques de « potentiel de contact » restent inévitables) en positionnant leur limite basse énergie à l'énergie du gap (ce qui revient à positionner l'origine des énergies au maximum de la bande de valence). Les spectres sont en énergie cinétique détectée. Aucune tension de collection n'est appliquée sur l'échantillon. Ces spectres sont normalisés de façon à ce que leur surface totale suive la variation du signal total de photoémission enregistré par ailleurs.

On ne commentera que les grandes tendances observables sur les spectres de la figure 5, en rappelant que leur interprétation est une affaire complexe si on ne dispose pas au minimum d'un bon calcul du spectre d'excitation dans le solide (ce qui n'est déjà pas un problème simple).

L'harmonique 13 produit un spectre très proche du spectre mesuré pour la partie imaginaire de la fonction diélectrique par Roberts et Walker [6] (pic au voisinage de 12 eV, épaulement basse énergie, et queue haute énergie). Dans le spectre avec l'harmonique 15 on retrouve ces traits, auxquels s'ajoute un épaulement du côté des hautes énergies. Du côté des harmoniques élevées on constate un aspect totalement différent. D'une part un pic très net apparaît du côté des basses énergies et d'autre part, on constate un déficit très net du côté des électrons de haute énergie (le trait s'arrête à l'énergie maximum que l'on atteint à partir du haut de la bande de valence par absorption d'un photon harmonique). Cette tendance naît dès l'harmonique 19 (l'harmonique 17 semblant un cas intermédiaire), et se renforce jusqu'à l'harmonique 27.

L'apparition d'électrons aux basses énergies peut être interprétée comme résultant du processus d'excitation de paires secondaires (multiplication des porteurs). On constate alors que ce processus ne devient réellement important que pour les harmoniques élevées, pour lesquelles l'énergie maximum des électrons atteint 30 à 35 eV, ce qui veut dire que l'énergie de ces électrons mesurée par rapport au bas de la bande de conduction est de l'ordre de l'énergie du plasmon dans le diamant (un modèle de Drude – non applicable pour un isolant – donne pour celui-ci 31.5 eV, mais un calcul de pertes d'énergie utilisant les théories ab-initio les plus récentes donne, pour la contribution des plasmons, une bande assez large s'étalant typiquement de 20 à 40 eV, ce qui correspond à la région grisée sur le plan de base de la figure 5).

Pour fixer les idées, nous avons fait figurer sur le panneau du fond le résultat d'un calcul RPA de pertes d'énergies, le pic centré à 40 eV correspondant aux pertes par émission de plasmon, et la contribution résiduelle à 15 eV aux pertes par excitations individuelles. Attention toutefois au fait que ce calcul est effectué pour des électrons dont l'énergie est grande devant l'énergie de plasmon, ce qui n'est pas le cas des électrons que nous excitons avec les harmoniques. Un calcul adapté à notre expérience reste à faire. Néanmoins, L'évolution des spectres UPS suggère que, si l'excitation de paires e-h individuelles est un processus possible, l'excitation d'un plasmon par les électrons d'énergie plus haute, suivie de la désintégration de ce plasmon en excitations individuelles est un processus plus efficace. Mais il ne s'agit que d'une hypothèse qui demande à être vérifiée à l'aide d'autres expériences, l'une d'entre elle étant évidemment une mesure de photoconductivité utilisant les harmoniques plus élevées. Par contre les spectres de la figure 5 ne suggèrent aucune explication évidente au comportement observé en photoconductivité.

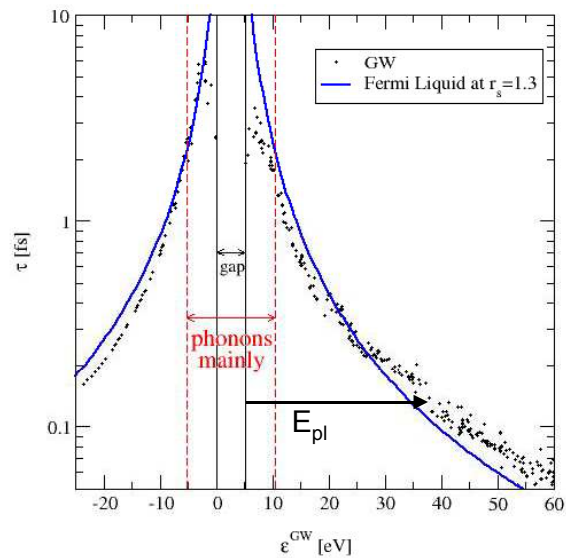

Figure 6 : calcul ab initio – dans l'approximation GW – de la durée de vie des porteurs libres ($E_{pl}$ : énergie RPA du plasmon)

Sur le plan théorique, à part les calculs évoqués ci-dessus (structure de bande LDA et spectres de pertes d'énergies), nous avons réalisé un premier calcul ab initio de la durée de vie des porteurs photoinjectés. Ce calcul consiste à calculer la partie imaginaire de l'énergie des « quasiparticules » qui sont la représentation théorique de ce que nous avons désigné sous le nom de porteurs, dans le cadre d'une approximation pour la self-énergie de celles ci connue sous le nom de « GW » [7]. Le résultat de ce calcul est montré sur la figure 6. Rappelons que l'approximation GW permet de prendre en compte les interactions électron-électron, individuelles et collectives, mais ni les interactions avec le réseau (phonons, défauts), ni les processus radiatifs, ni les interactions électron-trou (effets « excitoniques »). Ce calcul prend donc en compte des processus tels que l'excitation de paires secondaires (la multiplication des porteurs). Malgré son caractère préliminaire, les conclusions générales que nous en tirons ci dessous sont fiables.

Les résultats du calcul sont comparés à un modèle de type « liquide de Fermi » (LF) pour référence. Rappelons que ce modèle - qui prévoit une durée de vie en $1/(E-E_F)^2$ – est basé sur des arguments de type « espace de phase disponible » et sur une approximation de bande parabolique. le paramètre $r_S$ utilisé ne dépend que de la densité électronique du matériau et n'est pas ajustable. Ce modèle n'est en principe valide que pour des métaux simples. Pour notre calcul, le niveau de Fermi a été situé au milieu de la bande interdite. Il faut donc négliger la région du gap, où le calcul n'a pas de sens, mais aussi son voisinage immédiat, où la durée de vie est fortement affectée par le couplage avec les phonons. Compte tenu des ordres de grandeur auxquels on aboutit, les processus radiatifs (très lents) n'entrent pas en ligne de compte. Les énergies négatives correspondent aux (quasi)-trous, positives aux (quasi)-électrons, et on constate que pour des énergie faibles, le temps de vie calculé est légèrement inférieur à la valeur LF (mais en restant du même ordre). La situation change du côté électrons au voisinage de 20 eV (harmonique 13) où on remarque une augmentation de la durée de vie, qui rejoint la courbe LF, et passe au dessus. On remarque une seconde rupture à une énergie correspondant à l'énergie du plasmon (au passage de laquelle le temps de vie chute rapidement d'environ 30%. Remarquons que ces deux accidents interviennent aux énergies où nous avons constaté des comportements singuliers dans l'expérience (maximum dans la PC, apparition d'un fort pic de secondaires en UPS) : il ne s'agit pour le moment que de coïncidences, quoique dans le cas du plasmon, on obtient un tout cohérent : la durée de vie baisse car on ouvre un canal de désexcitation supplémentaire et, au moins dans les métaux, il est connu que les plasmons se désintègrent en excitations individuelles, ce qui est bien ce que semble indiquer l'expérience UPS.

En conclusion, nous avons étendu à l'utilisation d'harmoniques d'ordre élevé femtosecondes la méthode de photoconductivité induite par laser, développée dans le passé pour des lasers nano ou picoseconde, et nous l'avons appliquée à divers types de diamant. Cette méthode s'avère dans ce cas extrêmement sensible (probablement à cause de la grande mobilité des porteurs) permettant de détecter de l'ordre de $10^8$ porteurs, voire moins, et elle est particulièrement simple à mettre en œuvre (ne nécessitant pas le dépôt de contacts). Le comportement du signal de photoconductivité en fonction de l'ordre de l'harmonique ne suit pas ce que l'on déduirait du modèle souvent utilisé de multiplication des porteurs, en tout cas dans sa version simple : le rendement d'injection présente un maximum autour de H13, qui n'est pas pour le moment expliqué, mais pourrait être du à un effet de structure de bande. Les matériaux CVD et monocristallins montrent par ailleurs des comportement similaires (ce qui est loin d'être le cas quand on utilise une excitation dans le visible ou proche IR).

Nous avons mesuré les spectres UPS de l'échantillon monocristallin à l'aide des harmoniques 13 à 27. Le point le plus marquant est l'apparition d'un fort pic aux basses énergies quand on excite à des énergies voisines de l'énergie du plasmon, ce qui pourrait indiquer que le mécanisme de relaxation par émission de plasmons est particulièrement efficace.

Enfin nous avons effectué un calcul ab initio de la durée de vie des porteurs excités, dans le cadre de l'approximation GW, qui prends en compte les mécanismes d'excitations électroniques secondaires individuels et collectifs. on obtient en particulier une signature assez claire de l'effet des plasmons sur cette durée de vie.

Expérience comme calculs demandent à être raffinés, et en particulier il semble particulièrement intéressant d'observer le comportement de la photoconductivité induite par des harmoniques d'ordre élevé (typiquement entre 15 et 27) pour y rechercher une signature de l'effet des plasmons. On constatera aussi que la durée de vie calculée est de l'ordre de la femtoseconde ou moins, ce qui montre tout l'intérêt de la recherche sur la génération d'impulsions XUV attoseconde [8] si on souhaite un jour aborder de tels problèmes par des techniques pompe-sonde.

Les auteurs tiennent à signaler le support de la région Aquitaine, ainsi que du programme INTAS (N° 2001-458) et d'un PAI France-Russie (N° 04527 ZM) qui ont rendu possible la collaboration à l'origine de ces études.